\documentclass[iop,onecolumn]{emulateapj}

\usepackage{amsmath,amssymb,amstext,ulem}
\usepackage{natbib}
\usepackage{xcolor}
\usepackage{hyperref}

\newcommand{\kms}{km s$^{-1}$}
\newcommand{\rs}{{\rm R}_{\odot}}
\newcommand{\ndash}{\textendash}
\newcommand{\cor}{\mathrm{cor}}
\newcommand{\Ma}{M_{\rm A}}
\newcommand{\Mac}{M_{\rm A}^{\rm crt}}
\newcommand{\suma}{\mathlarger{\mathlarger{\sum}}}
\newcommand{\thetabn}{\theta_{\text{\tiny Bn}}}
\newcommand{\Bcor}{\vec{B}_{\cor}^{\up}}
\newcommand{\up}{\rm up}

\usepackage{isomath}
\renewcommand{\vec}[1]{{\mathbfit #1}}
\newcommand{\un}{\hat{\mathbfit{n}}}
\newcommand{\unr}{\hat{\mathbfit{r}}}

\newcommand{\unphi}{\hat{\mathbfit{\phi}}}
\usepackage{mathtools}
\DeclarePairedDelimiter\norm{\lVert}{\rVert} 
\usepackage{relsize}

\usepackage{setspace}
\shorttitle{Corrugated Shocks: A Discussion on the Predisposition to Particle Acceleration}
\shortauthors{P\'aez et al.}

\begin{document}

\title{corrugated features in coronal-mass-ejections-driven shocks: a discussion on the predisposition to particle acceleration}

\author{
A.~P\'aez \altaffilmark{1},
V.~Jatenco-Pereira \altaffilmark{1},
D.~Falceta-Gon\c{c}alves \altaffilmark{2},
\&  M.~Opher \altaffilmark{3}
}
\affil{
$^{1}$ Universidade de S\~ao Paulo, Instituto de Astronomia, Geof\'isica e Ci\^encias Atmosf\'ericas, Departamento de Astronomia, Rua do Mat\~ao 1226, S\~ao Paulo, SP, 05508-090, Brazil; andresspaez@usp.br \\
$^{2}$ Universidade de S\~ao Paulo, Escola de Artes, Ci\^encias e Humanidades, Rua Arlindo Bettio 1000, S\~ao Paulo, SP, 03828-000, Brazil\\
$^{3}$ Astronomy Department, Boston University, Boston, MA 02215, USA
}

\begin{abstract}
The study of the acceleration of particles is an essential element of research in the heliospheric science. Here, we discuss the predisposition to the particle acceleration around coronal mass ejections (CMEs)-driven shocks with corrugated wave-like features. We adopt these attributes on shocks formed from disturbances due to the bimodal solar wind, CME deflection, irregular CME expansion, and the ubiquitous fluctuations in the solar corona. In order to understand the role of a wavy shock in particle acceleration, we define three initial smooth shock morphologies each one associated with a fast CME. Using polar Gaussian profiles we model these shocks in the low corona. We establish the corrugated appearance on smooth shock by using combinations of wave-like functions that represent the disturbances from medium and CME piston. For both shock types, smooth and corrugated, we calculate the shock normal angles between the shock normal and the radial upstream coronal magnetic field in order to classify the quasi-parallel and quasi-perpendicular regions. We consider that corrugated shocks are predisposed to different process of particle acceleration due to irregular distributions of shock normal angles around of the shock. We suggest that disturbances due to CME irregular expansion may be a decisive factor in origin of particle acceleration. Finally, we regard that accepting these features on shocks may be the start point for investigating some questions in the sheath and shock, like downstream-jets, instabilities, shock thermalization, shock stability, and injection particle process.
\end{abstract}
\maketitle

\section{Introduction}
Coronal mass ejections (CMEs) are sporadic phenomena in solar surface that reconfigure notably the global coronal magnetic field \citep[e.g.,][]{2001JGR...10625141L, 2009ApJ...698L..51L}. The super-magnetosonic CMEs ($>800$ \kms) create a coronal shock wave in distances of $\sim 1.5 \, \rs$ \citep[e.g.,][]{2011ApJ...738..160M, 2014A&A...564A..47Z, 2016ApJ...833..216G}, evidenced through the radio Type II burst \citep{1950AuSRA...3..387W,1960PASJ...12..376U}, and Moreton waves \citep{1960AJ.....65U.494M,1960PASP...72..357M}. Together with the shock wave, the sheath structure is established by accumulating coronal plasma by the CME compression on medium. The shock and sheath generate conditions appropriated for particle acceleration \citep[e.g.,][]{2000JGR...10525079Z, 2005ApJ...622.1225M, 2013ApJ...778...43K}. In the shock wave, the particles are accelerated mainly through the diffusive shock acceleration process \citep[e.g.,][]{1978MNRAS.182..147B, 1978MNRAS.182..443B, 1978ApJ...221L..29B}. This type of particles are known as gradual solar energetic particles (electrons, protons, ions, hereafter SEPs, \citealt{1999SSRv...90..413R, 2013SSRv..175...53R}).

Some CMEs exhibit the shock signatures in the CME flanks or in the CME nose regions \citep[e.g.,][]{2009ApJ...693..267O}, consequently some events show the origin of SEPs in shock flanks \citep[e.g.,][]{2016ApJ...819..105K}, or at the shock nose \citep[e.g.,][]{1997ApJ...491..414R, 1999SSRv...90..413R}. In large SEPs events, fastest CMEs ($\sim 2000$ \kms) associated with the Ground Level Enhancement (GLE) events, the particle acceleration can occur from $\sim 2.0$ to $\sim 4.0\, \rs$, with average in $\sim 3.0 \, \rs$ \citep[e.g.,][]{2009ApJ...706..844R, 2012SSRv..171...23G}. The SEPs are accelerated in the shock supercritical regions. In these regions the downstream Alfv\'enic Mach number, $\Ma$, is larger than the critical Mach number, $\Mac$, for which the flows and sound velocities are equivalent \citep[e.g.,][]{1984JPlPh..32..429E}. According to the shock normal angle, $\thetabn$, between the shock normal and the upstream magnetic field, the supercritical shock can be manifested by two phases: the quasi-parallel ($0 \leq \thetabn \leq \pi/4$) and quasi-perpendicular ($\pi/4 \leq \thetabn \leq \pi/2$) (\citealt{2013pcss.book.....B}). Recently, \cite{2011ApJ...739L..64B} and \cite{2014ApJ...784..102B} analyzed the CME-driven shock occurred on 1999 June 11. In this shock the authors found supercritical and subcritical conditions in the shock nose and flanks, respectively, at distances of $\sim 2.6 \, \rs$. The authors affirm that their results are important to locate the zones of particle acceleration.

In last years some works have evidenced the importance of taking into account realistic properties of the solar corona and of the CMEs in the studies of acceleration and transport of the SEPs. For example, \cite{2005ApJ...622.1225M} showed the relevance of sheath structure and bimodal SW.  \cite{2011ApJ...739L..64B} showed the existence of supercritical regions in the shock front, \cite{2015ApJ...810...97S} analyzed the role of expansion and acceleration of a CME on particle acceleration. Recently, \cite{2017ApJ...836...36P} explored the dependence of the particle spectra with the initial CME radius.  \cite{2017ApJ...851...38K} showed that shocks can accelerate particles more efficiently when propagating in a streamer-like magnetic configuration than in the radial configuration.

In this work, we are interested in discussing the predisposition to particle acceleration in corrugated shocks fronts. For this, we compare two shock morphology types: smooth and corrugated. For both cases, we identify the quasi-parallel and quasi-perpendicular regions by calculating the angle $\thetabn$, throughout the shock angular width. The smooth shocks are modeled by assuming three fast CMEs each with different morphology and in different latitude locations \citep[e.g.,][]{2009ApJ...693..267O}. With these, we analyze the predisposition to particle acceleration in all latitudes. We construct the corrugated shocks imposing wave-like or undulations features from perturbations of surrounding media as, bimodal solar wind (SW), boundary wind \citep[e.g.,][]{2005ApJ...622.1225M, 2010ApJ...714L.128S, 2015ApJ...801..100S}, fluctuations of solar corona properties \citep[e.g.,][]{2005A&A...435.1123W, 2008ApJ...687.1355E, 2014A&A...564A..47Z}, and corrugated from deflection and irregular expansion of the CME piston, \citep[e.g.,][]{2009SoPh..259..143X, 2011SoPh..269..389S, 2013ApJ...775....5K, 2015ApJ...811L..36K, 2015ApJ...805..168K}, CME irregular expansion \citep[e.g.,][]{2011ApJ...728...41E}. Our calculation suggest some constraints between smooth shock morphology and the physical process defined through the shock normal angles. Our results for the corrugated shocks show the diversification of the quasi-parallel and quasi-perpendicular regions that directly affect the acceleration rate and the energy of the particles accelerated. Finally, we note that irregular CME expansions may be the most decisive factor in the predisposition of CME-driven shocks to accelerate particles.

This paper is organized as follows. In section~\ref{sec-model}, we construct the shocks morphologies in polar coordinates, and the corrugated shocks by imposing wave-like features on smooth shocks. In the section~\ref{sec-calc}, we identify the quasi-parallel and quasi-perpendicular regions through the shock width, in order to recognize some SEPs constraints in the upstream shock region. Finally, we show our discussion and conclusions in the section~\ref{sec-discuss}.

\section{Methodology}\label{sec-model}

In the same trend of incorporating realistic properties in studies of particle acceleration, our work focuses in discussing the predisposition of the wavy or corrugated shocks to the SEPs phenomenon. Few years ago, \cite{2015ApJ...812..119S} showed the CME-driven shock of 2011 June 7 between heliocentric distances from 2 to 12 $\rs$ and angular width of 110$^{\circ}$ by images of coronagraphs C2 and C3 of LASCO/SOHO. The authors show the shock fronts location with the irregular shock shape features, see Figure 5 in \cite{2015ApJ...812..119S}.  Similar irregular shock fronts were detected in the CME-driven shock of 1999 June 11 \citep{2011ApJ...739L..64B, 2013JAdR....4..287B, 2014ApJ...784..102B}. In order to understand the relevance of the corrugated shocks, we compare two shock morphology types: smooth and corrugated shocks. With this, our methodology is structured in two steps. First, in subsection~\ref{subsec-cor_sh}, we model the smooth shocks. Second, in subsection~\ref{subsec-sh_und}, we impose the undulations on top of a smooth shock, in order to mimic the disturbed shocked region due to external turbulence or subject to instabilities.

\begin{figure}
\centering
\includegraphics[width=0.9\textwidth]{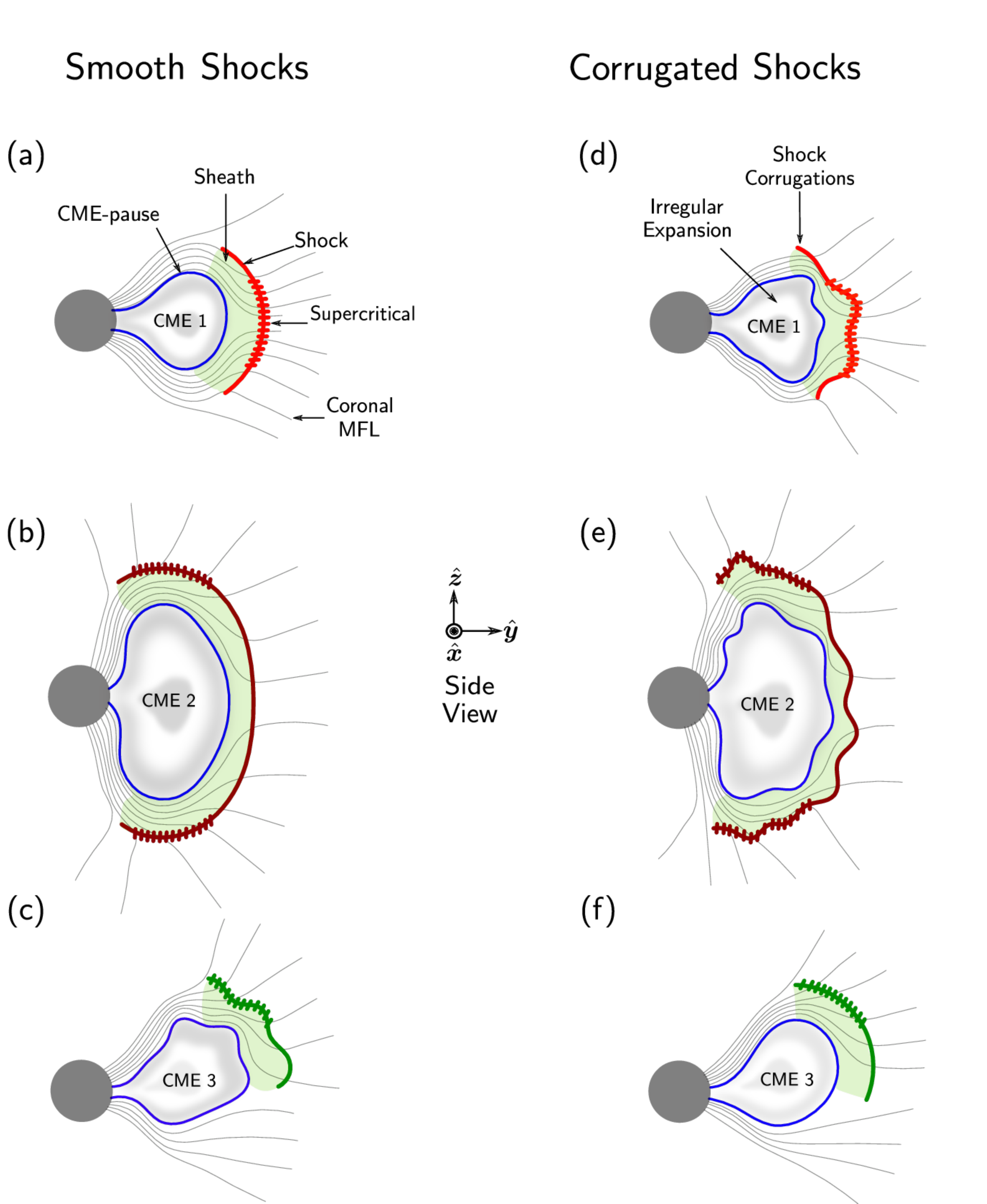}
\caption{Scheme of a meridional view of CME 1, CME 2, and CME 3 and their shocks. We show three different CMEs, structured in core; cavity; and frontal loop \citep{1985JGR....90..275I}. For all situations we considered a sheath structure (green shadow) formed behind the shocks. The CME-pause and coronal magnetic field lines (MFL) are indicated by the blue and black thin lines, respectively. Panels (a), (b), and (c) indicate the smooth shocks. For the three CME cases, the shock signatures (red, maroon, and green thick lines) are assumed in different latitude locations. The shock morphology preserves similar features to the events of 1999 September 11 (here assumed in the equator region), 1997 November 6, and 1998 June 4 studied in \cite{2009ApJ...693..267O}. Our interest with these three morphology types is to study all intervals of latitude. Panels (d), (e), and (f) show our propose of corrugated shocks. We model this type of shock by imposing wave-like features from bimodal SW \citep[e.g.,][]{2005ApJ...622.1225M}, CME deflection \citep[e.g.,][]{2013ApJ...775....5K, 2015ApJ...805..168K}, CME irregular expansions \citep[e.g.,][]{2011ApJ...728...41E}, and ubiquitous fluctuations of density and magnetic field of the solar corona \citep[e.g.,][]{2005A&A...435.1123W, 2008ApJ...687.1355E, 2014A&A...564A..47Z}. For the six cases, we calculate the shock normal angles throughout the shock width in order to provide a diagnostic general of the predisposition to particle acceleration in shock front. But particularly, we adopt the supercritical shock conditions at the convex regions (rounded, indicated by the red, maroon, and green transverse lines) assuming that these regions maintain high expansion velocity  \citep[e.g.,][]{2011ApJ...739L..64B, 2014ApJ...784..102B}.}\label{fig-cmes}
\end{figure}

Figure~\ref{fig-cmes} illustrates the side view configurations of the six CME-driven shocks analyzed in this paper. We consider three CMEs: CME 1, CME 2, and CME 3 and their shocks waves in different latitudinal locations (red, maroon, and green thick lines, these color features are conserved through the paper). Panels (a), (b), and (c) show the smooth CMEs and theirs shocks. These shock morphologies have similar features to the events of 1999 September 11 (here assumed in the equator region), 1997 November 6, and 1998 June 4 analyzed in \cite{2009ApJ...693..267O}. Our interest in these shocks morphology shock morphologies is to study all the interval of latitude considering the shock 1 at CME 1 nose, shock 2 at CME 2 flanks, and shock 3 at the intermediate latitude. For these CMEs, we consider high velocities $\gtrsim 1500$ km s$^{-1}$, and cone-like structure in three part-structure: core, cavity and frontal loop \citep{1985JGR....90..275I}. The CME-pause and coronal magnetic field lines (MFL) are indicated by blue and black thin lines, respectively. In all cases, the sheath structure (green shadow) is assumed and the CME magnetic reconnection is neglected. Panels (d), (e), and (f) illustrate our model of corrugated CME pistons and shocks. For the six cases, we analyze the shock width in order to provide a general diagnostic of shock normal angles. But particularly, we adopt the supercritical shocks conditions at the convex regions indicated by the red, maroon, and green transverse lines. We consider that these regions have high Mach number, due to its faster expansion velocity, relative to the others shock regions \citep[e.g.,][]{2011ApJ...739L..64B, 2014ApJ...784..102B}.

\subsection{Coronal smooth shocks model}\label{subsec-cor_sh}

We model the smooth shock surface by polar Gaussian plots, $S_m(\phi)$, as function of latitude coordinate, $\phi$ (e.g., \cite{2009ApJ...702..901W}, \cite{2010ApJ...715.1524W}, and \cite{2011ApJ...729...70W} for CMEs). The subscript $m$ with values $m = 1, 2, 3$ identify the shock morphology associated to each CME, see Figure~\ref{fig-cmes}. The shocks locations are adjusted close to $\sim 3.0 \, \rs$ according to the coronal distances of shock formation at $\sim 1.5 \, \rs$ \citep[e.g.,][]{2011ApJ...738..160M, 2014A&A...564A..47Z, 2016ApJ...833..216G}, particle acceleration onset between $\sim 2.0$ to $\sim 4.0 \, \rs$ \citep[e.g.,][]{2009ApJ...706..844R, 2012SSRv..171...23G}, and supercritical shock detections $\sim 2.6 \, \rs$ \citep[e.g.,][]{2011ApJ...739L..64B, 2014ApJ...784..102B}.

\begin{figure*}
\centering
\includegraphics[width=0.7\textwidth]{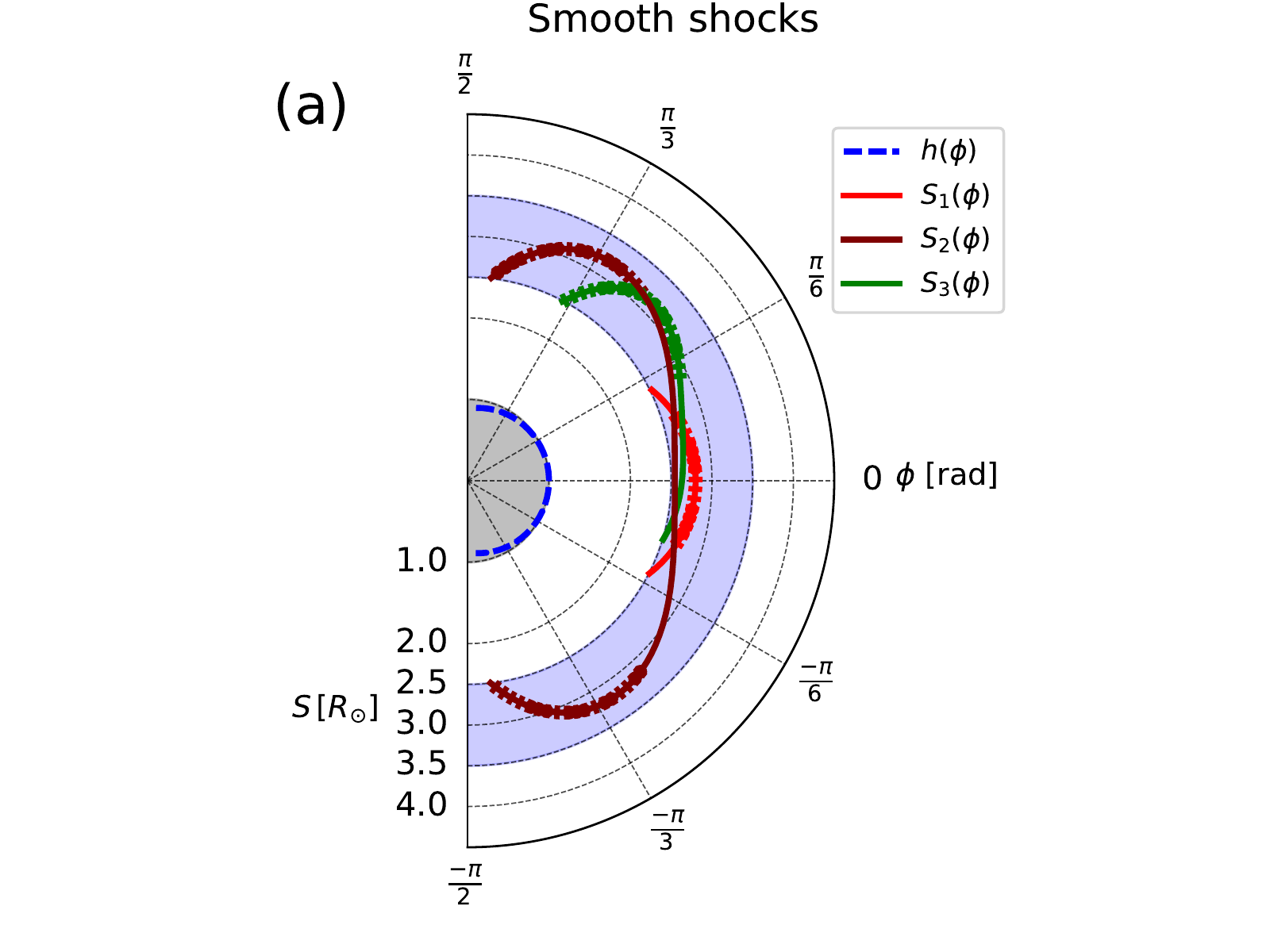} \\ \includegraphics[width=0.7\textwidth]{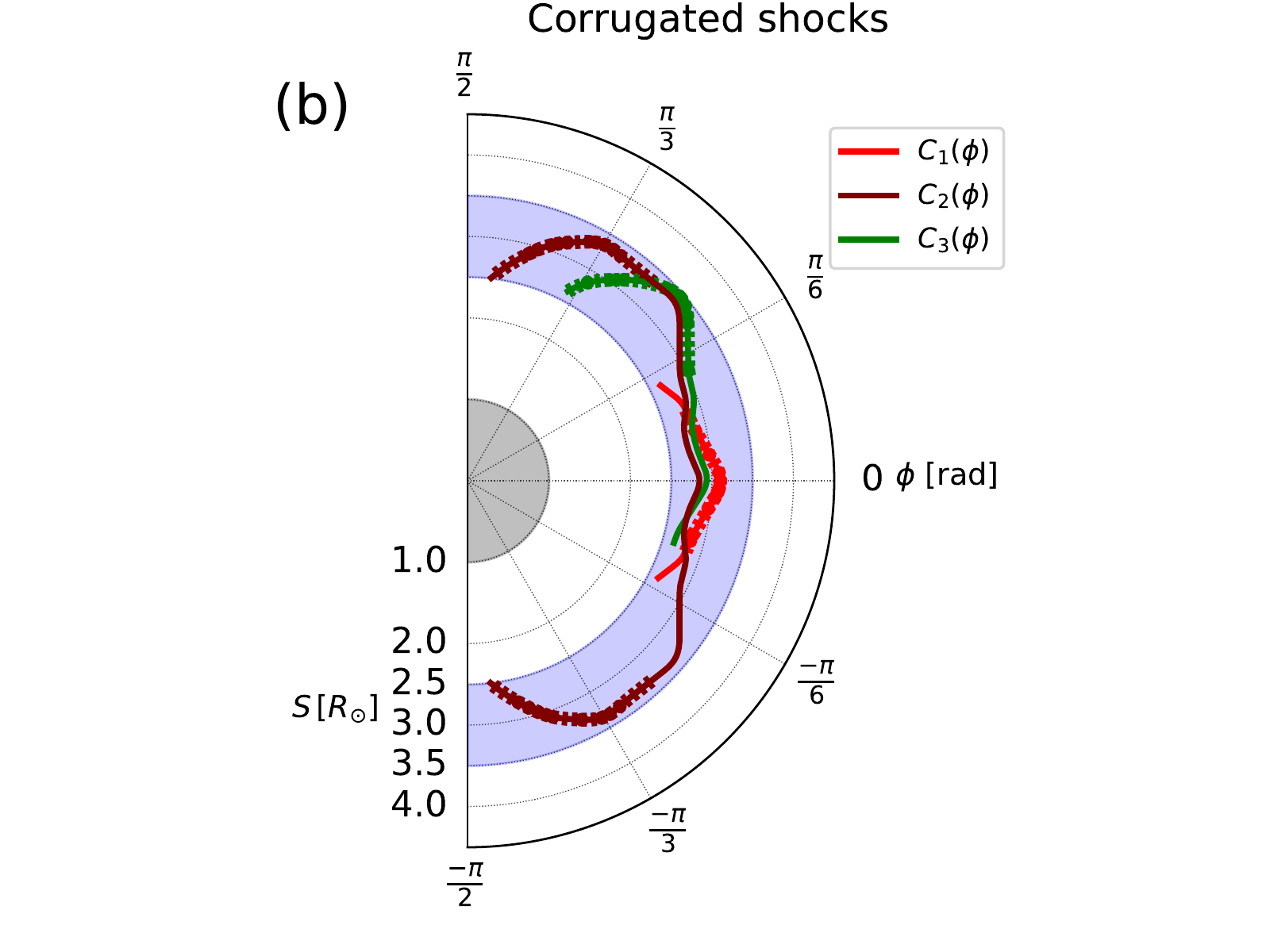}
\caption{Reconstruction of the shocks presented in Figure~\ref{fig-cmes}. Panel~(a) shows the function $h(\phi)$ (blue dashed line) Equation~\eqref{eq-hphi}, together with the smooth shocks $S_1(\phi)$ (red line), $S_2(\phi)$ (maroon line), and $S_3(\phi)$ (green line), Equations~\eqref{eq-shock1}, \eqref{eq-shock2}, and \eqref{eq-shock3}, respectively. In panel~(b) we show the corrugated shocks $C_1(\phi)$ (red line), $C_2(\phi)$ (maroon line) and $C_3(\phi)$ (green line) Equation~\eqref{eq-shock_und}. The blue shadow between 2.5 to 3.5 $\rs$ corresponds to the region where the shocks are analyzed. In this interval we considerate the shock formation ($\sim 1.5 \, \rs$, e.g., \citealt{2011ApJ...738..160M, 2014A&A...564A..47Z, 2016ApJ...833..216G}) and particle acceleration ($\sim 3.0 \, \rs$, e.g., \citealt{2009ApJ...706..844R, 2012SSRv..171...23G}). For both shock types we identify the quasi-parallel and quasi-perpendicular regions. The red, maroon, and green transverse lines at convex (round, regions with high expansion velocity) regions indicate the locations where we assumed the supercritical shock conditions \citep[e.g.,][]{2011ApJ...739L..64B, 2014ApJ...784..102B}. The gray half-circle represents the Sun.}\label{plot-shocks}
\end{figure*}

We build the smooth shocks, $S_m(\phi)$, from an initial parabolic-like shape profile,
\begin{equation}\label{eq-hphi}
h(\phi) = \exp{ \bigg(- \frac{\phi^2}{2} \bigg) }.
\end{equation}
From CMEs observation it is possible to consider that these parabolic-like shape can be more realistic than circular profiles due to the nonuniform ejecta driver. Close to the SEPs onset ($\sim 3.0 \, \rs$), we define the first shock surface, $S_1(\phi)$ as multiple of $h(\phi)$,
\begin{equation}\label{eq-shock1}
S_1(\phi) = 2.8 \, h(\phi).
\end{equation}
This shock is the simplest CME-driven shock morphology. 
$S_1(\phi)$ has been chosen in order to mimic the CME event of 1999 September 11 studied in \cite{2009ApJ...693..267O} (here adapted for the equator region).

The shocks $S_2(\phi)$ and $S_3(\phi)$, Figure~\ref{fig-cmes}(b) and~(c), are more complex than $S_1(\phi)$. In order to establish these shocks, we start from $h(\phi)$ in combination with an auxiliary function structured as the sum of polar Gaussian functions,
\begin{equation}\label{eq-pm_function}
p_m(\phi) = \suma\limits_{i=-9}^{9} a_i \exp \Bigg[ -b_i \Bigg( \phi- \frac{i \, \pi}{20} \Bigg)^2 \Bigg].
\end{equation}
The mathematical flexibility of the $p_m(\phi)$ function through its parameters of amplitude ($a_i$), width ($b_i$), and locations ($i\pi/20$), and the positive range ($\geq 0$) of the Gaussian profiles, allow $S_2(\phi)$ and $S_3(\phi)$ to be written as a combination of $h(\phi)$ and $p_m(\phi)$ as,
\begin{equation}\label{eq-shock2}
S_2(\phi) =  1.5 \, h(\phi) + p_2(\phi),
\end{equation}
and 
\begin{equation}\label{eq-shock3}
S_3(\phi) = 2.3 \, h(\phi) + p_3(\phi).
\end{equation}
The 1.5 and 2.3 multiples of $h(\phi)$ function are taken assuming the shock close to $3.0 \, \rs$. Table~\ref{tab-constants} shows the positive constants $a_i$ and $b_i$. These are adjusted in order to the shocks $S_2(\phi)$ and $S_3(\phi)$ to be latitudinally symmetric and mimic the shock shape of 1999 May 27, and 1998 June 4 \citep{2009ApJ...693..267O}, respectively.

Figure~\ref{plot-shocks}(a) shows the profiles, $h(\phi)$ (blue dashed line), and the shock functions $S_1(\phi)$ (red line), $S_2(\phi)$ (maroon line), and $S_3(\phi)$ (green line). Panel (b) shows the plots of the corrugated modeled shocks in subsection~\ref{subsec-sh_und}. The red, maroon, and green transverse lines at the shocks convex regions indicate our initials zones with supercritical shock conditions. In plots the blue shadow from 2.5 $\rs$ to 3.5 $\rs$ indicate the interval where we analyze the shocks. In Section~\ref{sec-calc} we expand our motivation in this interval.

\subsection{Coronal corrugated shocks model}\label{subsec-sh_und}

During the early evolution stages of the CME at the low corona suffers several disturbances from medium that causes non-uniform expansion and deflection. We consider that these disturbances impose wave-like features on the CME piston, and consequently on the shock front. These perturbations to the shape and velocity of the shock front are known as corrugation instability (e.g., \citealt{1964PhFl....7..700G} and \citealt{1987flme.book.....L}). In our work, we scale the factors that can disturb the shock: (\textit{i}) large-scale amplitudes from SW and CME deflection, (\textit{ii}) medium-scale from CME properties and configuration and (\textit{iii}) small-scale from fluctuations of solar corona properties.

The SW effect on shock may be due to the difference of velocities between the slow and fast wind \citep[e.g.,][]{2005ApJ...622.1225M, 2010ApJ...714L.128S}, as also to its boundary wind \citep{2015ApJ...801..100S}. Besides the CME deflection possibly may impose the decentralization of the shock with respect to the CME, or to the shock deformation comparable to the showed in \cite{2012ApJ...755...43W} and \cite{2013ApJ...778...43K}. It is known that CME deflection is established from 2.0 $\rs$ \citep{2015ApJ...811L..36K} as a consequence of the magnetic forces of tension and pressure gradient, together to the non-radial drag force of the SW background allow latitudinal and longitudinal CME deflections \citep{2013ApJ...775....5K, 2015ApJ...805..168K}. Thus, the SW and CME deflection effects on ejecta may be large angular undulations, possibly imposed by the decentralization of the shock with respect to the CME, or to the shock deformation similar to the showed in \cite{2012ApJ...755...43W} and \cite{2013ApJ...778...43K}.

Certainly the shock preserve the most relevant features of the ejecta piston, which can be due to magnetic configurations and irregular CME expansions. We classify these as medium-scale perturbations. \cite{2011ApJ...728...41E} show how the CME-pause (that represent the boundary in equilibrium between CME and shocked coronal plasma) can maintain an irregular shape. The authors explained it in two different ways. First, they affirm that it may be the result of magnetic field configurations relative to CME and the coronal and active region magnetic fields. Second, they suggest that it may be due to the deflected flows in the downstream region, similarly to the heliosheath \citep[e.g.,][]{2007Sci...316..875O, 2009Natur.462.1036O}. The CMEs can be affected by their irregular expansion due to the imbalance among internal magnetic and gas pressure, and external pressures of solar corona. It may allow some CME regions to expand more rapidly than others by dynamical pressure effects.

Besides the small-scale factors on shocks may be the ubiquitous irregularities of the density and magnetic field in the solar corona that give rise to fluctuations in the Alfv\'en velocity, therefore the shock front may be modified \citep[e.g.,][]{2005A&A...435.1123W, 2008ApJ...687.1355E, 2014A&A...564A..47Z}. In this paper we do not take into account the CMEs rotations \citep[e.g.,][]{2009ApJ...697.1918L, 2009ApJ...705..426Y} in order of guarantee the shock coplanarity hypothesis \citep[e.g.,][]{2013pcss.book.....B}. Also we neglected the CMEs interactions (e.g., \citealt{2017SoPh..292...64L} and references therein), but we highlight that these may affect substantially the shock fronts as well. Also we consider that shock disturbances may be consequence of the CME driver evolution, one reason could be internal reconnection as suggested by \cite{2014PhRvL.113c1101F}.

\begin{table}
\caption{Values of the constants $a_i$ ($\rs$), and $b_i$ (dimensionless) of the $p_m(\phi)$ function, Equation~\eqref{eq-pm_function}, in order to contructed the symetric shocks 2 and 3, Equations~\eqref{eq-shock2} and \eqref{eq-shock3}, repectively.}\label{tab-constants}
\begin{center}
\begin{tabular}{c||c|c|c|c|c|c|c|c|c|c}
\hline   \hline
i & 0 & 1 & 2 & 3 & 4 & 5 & 6 & 7 & 8 & 9 		\\
\hline
\multicolumn{11}{c}{shock 2}   \\
\hline    
$a_i$ & 0.0 & 0.0 & 0.0 & 0.0 & 0.0 & 0.0 & 0.6 & 0.9 & 0.7 & 0.2	 \\ \hline
$b_i$ & 0.0 & 0.0 & 0.0 & 0.0 & 0.0 & 0.0 & 1.2 & 2.5 & 0.6 & 3.5	  \\ \hline
\hline 
\multicolumn{11}{c}{shock 3}   \\
\hline    
$a_i$ & 0.0 & 0.0 & 0.0 & 0.3 & 0.2 & 0.4 & 0.7 & 0.0 & 0.0 & 0.0		\\\hline
$b_i$ & 0.0 & 0.0 & 0.0 & 0.0 & 7.0 & 5.0 & 7.0 & 0.0 & 0.0 & 0.0   	\\ \hline
\hline
\end{tabular}
\end{center}
\end{table}

The complexity of the CMEs, solar corona, and SW  allow that their perturbations in shock to be random. As initial approximation in this work we model these disturbances through wave functions. Mathematically we define the corrugated shocks, $C_m(\phi)$, imposing tenuous undulations on smooth shocks $S_m(\phi)$ by addition of a supplementary function $k(\phi)$, 
\begin{equation}\label{eq-shock_und}
C_m(\phi) = S_m(\phi) + k(\phi).
\end{equation}
The $k(\phi)$ is defined as the sum of smaller corrugated functions,
\begin{align}\label{eq-kphi}
k(\phi) = k_1(\phi) + k_2(\phi) + k_3(\phi),
\end{align}
where,
\begin{align}\label{eq-kphi_1}
k_1(\phi) = \suma\limits_{i=-1}^{1} 0.15 \exp{ \Bigg[ -30 \, \Bigg( \phi- \frac{2 \, i \, \pi}{9} \Bigg)^2 \Bigg] },
\end{align}
\begin{align}\label{eq-kphi_2}
k_2(\phi) = \suma\limits_{i=-3}^{3} 0.1 \exp{ \Bigg[ -60 \, \Bigg( \phi- \frac{i \, \pi}{9} \Bigg)^2 \Bigg] },
\end{align}
and,
\begin{align}\label{eq-kphi_3}
k_3(\phi) = \suma\limits_{i=-7}^{7} 0.05 \exp{ \Bigg[ -120 \, \Bigg( \phi- \frac{i \, \pi}{18} \Bigg)^2 \Bigg] }.
\end{align}

\begin{figure}
\centering
\includegraphics[width=0.7\textwidth]{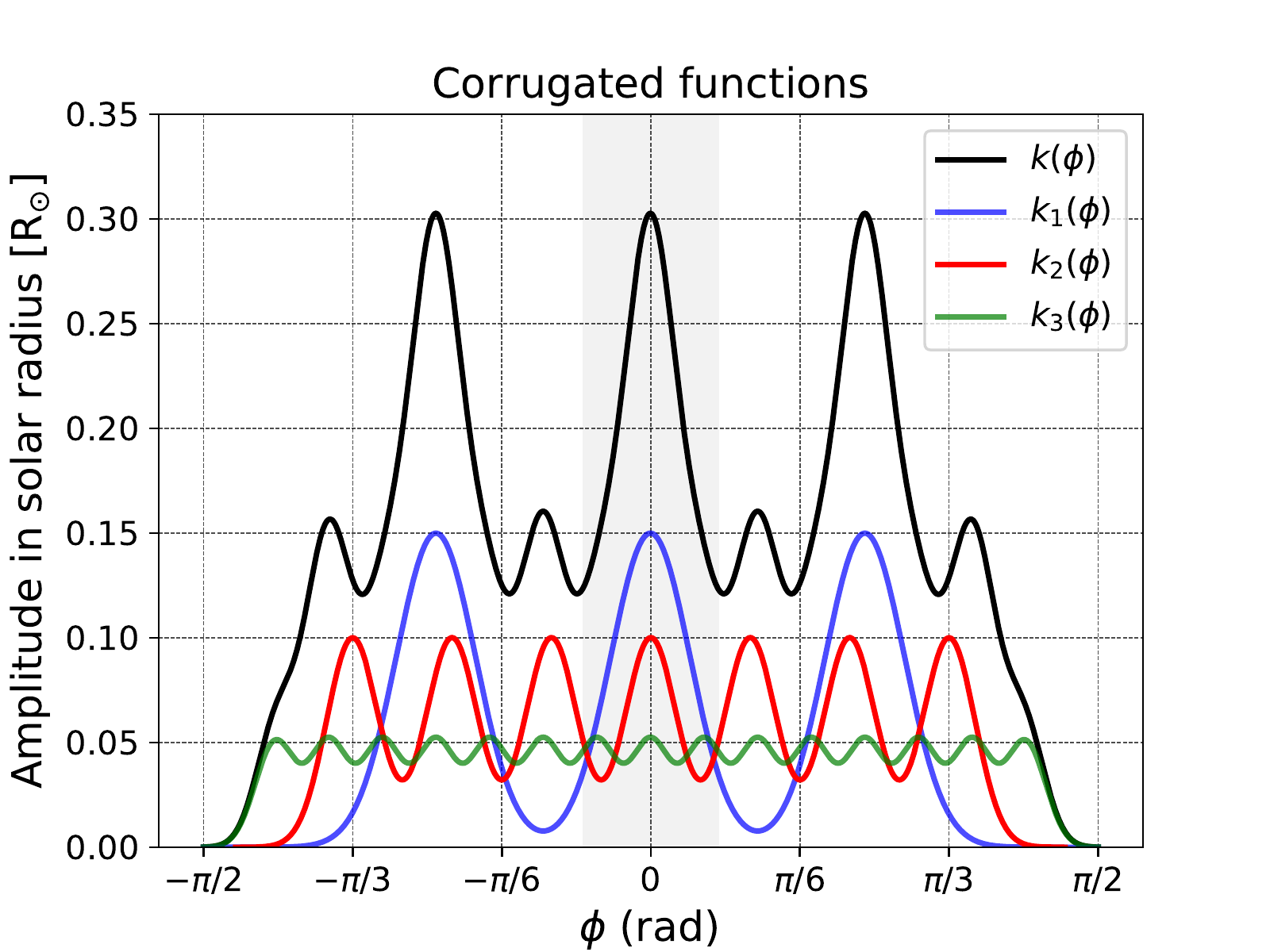}
\caption{Corrugated function $k(\phi)$ ($\rs$), Equation~\eqref{eq-kphi} (black line), and its contributions $k_1(\phi)$, Equation~\eqref{eq-kphi_1} (blue line), $k_2(\phi)$, Equation~\eqref{eq-kphi_2} (red line); and $k_3(\phi)$, Equation~\eqref{eq-kphi_3} (green line) in function of polar angle coordinate, $\phi$. With $k_1(\phi) > k_2(\phi) > k_3(\phi)$ amplitudes, we scale the effect of the bimodal SW, CME irregular expansion and fluctuations of solar corona as large, medium and small scales for $\sim 3.0 \, \rs$ distances. The $k(\phi)$ function is imposed on smooth shocks $S_m(\phi)$, Equations~\eqref{eq-shock1}, \eqref{eq-shock2}, and~\eqref{eq-shock3}, to construct the corrugated shock, $C_m(\phi)$, Equation~\eqref{eq-shock_und}. The gray shadow indicate the amplitude of 0.3 $\rs$ and angular width $\sim \pi/6$ rad ($\sim 30^{\circ}$) of the larger undulations. }\label{plot-kphi}
\end{figure}

Figure~\ref{plot-kphi} shows the functions $k(\phi)$ (black line), $k_1(\phi)$ (blue line), $k_2(\phi)$ (red line), and the $k_3(\phi)$ (green line). The $k_1(\phi)$ function, Equation~\eqref{eq-kphi_1}, is intentionally structured with three wave crests, in order to represent the disturbances on shock caused by the SW interfaces together with the fast and slow SW. With $k_2(\phi)$ function, Equation~\eqref{eq-kphi_2}, we represent the disturbances due to the irregular CME expansion. In $k_3(\phi)$ function, Equation~\eqref{eq-kphi_3}, we take into account the minor disturbances in the shock induced by fluctuations in the density, Alfv\'en velocity or magnetic field strength of the solar corona. The amplitudes of the $k_1(\phi)$ ($\sim 15$\% $\rs$), $k_2(\phi)$ ($\sim 10$\% $\rs$), and $k_3(\phi)$ ($\sim 5$\% $\rs$) compose a corrugated function $k(\phi)$ with maximum amplitude of 30\% $\rs$ and crest angular width of the $\sim \pi/6$ rad ($\sim 30^{\circ}$) indicated by the gray shadow. With $k_1(\phi) > k_2(\phi) > k_3(\phi)$ amplitudes, we scale the effect of the bimodal SW, CME irregular expansion and fluctuations of solar corona as large, medium and small scales for $\sim 3.0 \, \rs$ distances. In Figure~\ref{plot-shocks}(b) we show the corrugated shocks $C_1(\phi)$ (red line), $C_2(\phi)$ (maroon line) and $C_3(\phi)$ (green line).

\section{Calculation}\label{sec-calc}

In order to understand the predisposition to particle acceleration in undulated shock fronts, we compare the distributions of quasi-parallel and quasi-perpendicular regions along of the smooth and corrugated shocks shown in Figures~\ref{fig-cmes} and~\ref{plot-shocks}.  We identify these regions by calculating the shock normal angle, $\thetabn$, between the shock normal, $\un$, and the upstream coronal magnetic field, $\vec{B}_{\cor}^{\up}$. We assume a steady global coronal magnetic field with open magnetic field lines in the polar regions and closed field lines at low latitudes of the equator besides we consider the bimodal structure of the SW \citep[e.g.,][]{2004JGRA..109.1102M}. For distances larger than the source surface radius, in our calculation at 2.5 $\rs$ \citep[e.g.,][]{1969SoPh....9..131A}, we consider the $\Bcor$ disposed radially in the upstream region (e.g., \citealt{2014ApJ...784..102B} and references therein). The supercritical shock regions in our models are assumed in the radial configuration of the $\Bcor$, i.e., $>2.5 \, \rs$. In this range, we do not take into account the closed field structures that are ubiquitous in the solar corona for distances $<2.5 \, \rs$. We point that in an eventual interaction of the shock wave with the closed magnetic field lines the $\thetabn$ values may change to values close to $\sim \pi/2$, this due to the orthogonality between normal shock and closed magnetic field lines. In this interaction it is possible to assume the curvature of the shock larger than the curvature of the closed field lines. For this condition, \cite{2016ApJ...821...32K} suggest that particles (electrons in \citeauthor{2016ApJ...821...32K} study) are swept by shock toward the shock flanks where are accelerated.

In Figures~\ref{fig-plane_shock}(a) and~(b) we compare the smooth and corrugated shocks (red lines) and CMEs piston (gray shadow), CME-pause (blue line), sheath or downstream (green shadow), upstream (yellow shadow), and the angles $\thetabn$, between $\un$ (blue arrows), and $\vec{B}_{\cor}^{\up}$ (black thin arrows). In the corrugated shock, we characterize the undulations with a lower amplitude than that in the CME piston. In order to show the contrast between two shocks types, we use the parabolic shock morphology, i.e., like the shock 1. The cartoon illustrate some features in shocks. \textit{(i)} The magnetic deflection between the downstream magnetic field lines that drape the CME and the upstream radial magnetic field lines for the $>2.5 \, \rs$ \citep[e.g.,][]{2010ApJ...720..130B, 2011ApJ...739L..64B, 2014ApJ...784..102B, 2015ApJ...809...58B}, (see Figure~7 in \citealt{2014ApJ...784..102B}). 
This deflection is also mentioned in \cite{2011A&A...527A..46L} as the rotation of the magnetic field in the downstream. This deflection is explained by the shock transit on the magnetic field that drape the CME \cite[e.g.,][]{2015ApJ...809...58B}. Additionally, some MHD numerical shock studies show the magnetic field amplification in downstream region and parallel upstream magnetic field disposition explained through that perpendicular downstream magnetic field is not advected \citep[e.g.,][]{2012MNRAS.423.1562F, 2015MNRAS.446..104R}. 
\textit{(ii)} The turbulent behavior in the sheath is indicated by means of curved black arrows in the sheath region \citep[e.g.,][]{2005ApJ...622.1225M}. \textit{(iii)} The magenta filamentary structures in the upstream regions illustrate the quasi-parallel (Q{\ndash}$\parallel$) regions where the particles can be removed more easily. These structures may be similar to the field aligned structures in the magnetospheric shock in the Q{\ndash}$\parallel$ regions \citep[e.g.,][]{2014JGRA..119.2593O}. The distribution of the quasi-parallel (magenta upstream filamentary structures) and quasi-perpendicular regions illustrate our results shown in Figure~\ref{plot-thetabn_angles} for both shocks types.

\subsection{Shock normal angles calculation}\label{subsec-quasipp}

We calculated the angles, $\thetabn$, between the normal of the shock, $\un$, 
and radial magnetic field, $\Bcor = \norm{\vec{B}_{\cor}^{\up}} \, \unr$, along the shock,
\begin{equation}\label{eq-coseno}
\cos \thetabn = \frac{ \un \cdot \vec{B}_{\cor}^{\up} }{ \norm{\vec{B}_{\cor}^{\up}} }  = \un \cdot \unr.
\end{equation}
We introduce $\un$, by rotating the tangential vector, $\vec{\tau}$, to the shock. 
This process consists of three steps: first, we define the shock surface, $S(\phi)$ (also corrugated shocks, $C_m(\phi)$ Equation~\eqref{eq-shock_und}), as a parametric function ($\big \langle \, \big \rangle$) of $\phi$, i.e.,
\begin{equation}
S(\phi) = \Big \langle \phi, S \Big \rangle.
\end{equation}
Second, we calculate the tangential vector, in polar components ($\unphi, \unr$),
\begin{equation}
\vec{\tau} = \unphi + S_{\phi} \, \unr,
\end{equation}
with $S_{\phi} = \frac{dS}{d\phi}$. Third, we rotate $\pi/2$ rad, in order to find the unitary normal vector,\begin{equation}\label{eq-normal}
\un = \frac{ -S_{\phi} \, \unphi \, + \, \unr }{ \left( S_{\phi}^2 +1 \right)^{\frac{1}{2}} } .
\end{equation}
With Equations~\eqref{eq-coseno} and \eqref{eq-normal}, we find the angle $\thetabn$,
\begin{equation}\label{eq-thetabn}
\thetabn = \arccos \left( S_{\phi}^2 +1 \right)^{-\frac{1}{2}} \, \text{rad}.
\end{equation}

Figure~\ref{plot-thetabn_angles} shows the shock normal angles for the smooth, $\thetabn^{\rm S_m}(\phi)$, and the corrugated $\thetabn^{\rm C_m}(\phi)$ shocks, with $ m = 1, 2, 3 $, shown in Figure~\ref{plot-shocks}. 
Panel (a) shows the $\thetabn$ values for the three smooth shocks, while panels (b), (c) and (d) show separately the $\thetabn^{\rm S_m}(\phi)$ (black dotted line), and $\thetabn^{\rm C_m}(\phi)$ (colored continuous line) plots. 
The color plots are associated with the colored features presented in this paper, i.e., the shocks 1, 2, 3 in red, maroon and green, respectively. The gray and white background shadows indicate the quasi-parallel (Q{\ndash}$\parallel$, $0 \leq \thetabn \leq \pi/4$) and quasi-perpendicular (Q{\ndash}$\perp$, $\pi/4 \leq \thetabn \leq \pi/2$) ranges of $\thetabn$.

In Figure~\ref{plot-thetabn_angles}(a), the angle $\thetabn^{\rm S_1}(\phi)$ (red line) oscillates from quasi-perpendicular to parallel angles between flanks to the nose of the $S_1(\phi)$ (red line in Figure~\ref{plot-shocks}). In this case there is only one point where the shock is completely parallel, i.e., $\thetabn = 0$ rad. The $\thetabn^{\rm S_2}(\phi)$ (maroon line), is more complex in comparation with $\thetabn^{S_1}(\phi)$. $\thetabn^{S_2}$ oscillate from quasi-perpendicular in shock flanks to the quasi-parallel to quasi-perpendicular values close to $\pm \pi/6$. The $\thetabn^{S_3}(\phi)$ (green line) can be considered an intermediate morphology, between those of $S_1(\phi)$ and $S_2(\phi)$, i.e., like deflection of the $S_1(\phi)$ or $S_2(\phi)$. Is interesting that $\thetabn^{\rm S_3}(\phi)$ profile can be interpreted like translational to the $\thetabn^{\rm S_1}(\phi)$ or $\thetabn^{\rm S_2}(\phi)$. For the three shock cases we do not find $\thetabn$ angles of perpendicular cases, i.e., $\thetabn = \pi/2$ rad, between normal shock and radial $\Bcor$. But, we comment that $ \thetabn $ can be $\sim \pi / 2 $ in situations where the CME-driven shock preserves like-pointed sharp features, possibly due to fast CME expansions.

\begin{figure}
\centering
\includegraphics[width=0.6\textwidth]{f4.pdf}
\caption{Schematic comparative between smooth, panel (a), and corrugated shocks, panel (b).
In order to show the contrast between two shocks types, we use the parabolic shock morphology, i.e., like the shock 1. 
We illustrate the differences between CME piston (gray shadow), CME-pause (blue line), downstream or sheath (green shadow), upstream (yellow shadow) of the shock (red lines), and distribution of the shock normal vectors ($\un$, blue arrows). For both cases we assume the coronal magnetic field lines (MFL) (black thin arrows) turbulent disposed in the sheath region \citep[e.g.,][]{2005ApJ...622.1225M}, and radially in the upstream region \citep[e.g.,][]{2010ApJ...720..130B, 2011ApJ...739L..64B, 2014ApJ...784..102B, 2015ApJ...809...58B}. We classify the quasi-parallel (Q{\ndash}$\parallel$) and quasi-perpendicular (Q{\ndash}$\perp$) regions by calculation of the shock normal angle, $\thetabn$, between shock normal, $\un$, and the radial upstream coronal magnetic field, ($\vec{B}_{\text{\tiny{cor}}}^{\up}$). Our results for the smooth and corrugated shock in Figure~\ref{plot-thetabn_angles} are illustrated by irregular distributions filamentary structures (magenta structures) in the Q{\ndash}$\parallel$ usptream regions \citep[e.g.,][]{2014JGRA..119.2593O}.}\label{fig-plane_shock}
\end{figure}

We find that any shock shape may be interpreted as a composition of convex (outward curvature) and concave (inward curvature) shapes as $S_1(\phi)$. Consequently its $\thetabn^{\rm S_m}(\phi)$ profiles also may be a composition of $\thetabn^{\rm S_1}(\phi)$. This can be inferred by comparing shocks 1 and 2. The $S_1(\phi)$ profile shows a convex profile and consequently its $\thetabn^{\rm S_1}(\phi)$ exhibits a ``V'' shape. $S_2(\phi)$ can be interpreted as two convex regions similar to $S_1(\phi)$ at extremes. The second finding, is the interesting behavior between shock 1 and shock 3. Figure~\ref{plot-shocks}(a) shows $S_3(\phi)$ as a deflection of the $S_1(\phi)$. This deflection effect may be considered in Figure~\ref{plot-thetabn_angles}(a) as an angular translation of $\thetabn^{S_3}(\phi)$ with respect to $\thetabn^{S_1}(\phi)$. $\thetabn^{S_3}(\phi)$ shows differences for $\phi < \pi/6$ rad due to the variations in amplitude and shock angular width of $S_3(\phi)$ with respect to $S_1(\phi)$.

Figures~\ref{plot-thetabn_angles}(b), (c)~and~(d) show the effects of the wave-like features of $k(\phi)$, Equation~\eqref{eq-kphi}, on the smooth shocks by comparing the $\thetabn^{\rm C_m}(\phi)$ (colored continuous lines) and $\thetabn^{\rm S_m}(\phi)$ (black dotted lines). The $C_m(\phi)$ and $\thetabn^{\rm C_m}(\phi)$ do not show constraints similar to the previous for the smooth shock. Besides the small amplitude of $k(\phi)$, i.e., $\leq 0.3 \, \rs$, affect drastically to the $\thetabn$ values. The most visible difference between $\thetabn^{\rm S_m}(\phi)$ and $\thetabn^{\rm C_m}(\phi)$ are the consecutive changes of $\thetabn$ and extreme values in $\thetabn^{\rm C_m}(\phi)$. 
The undulations notably multiply the parallel angles, i.e., $\thetabn \approx 0$ rad, along the shocks width, but do not allow the existence of the perpendicular angles, the last similar to the smooth shocks. The quasi-parallel and quasi-perpendicular regions maintain different behaviors that can show different rates of acceleration. The fast reconnection rate in quasi-perpendicular regions is a  consequence of a small coefficient of diffusion. Besides, it is known that short acceleration times, are related to high energies. Therefore, the quasi-perpendicular regions in the shock are expected to be those with particles with high energy \citep{2005ApJ...624..765G}. While in the quasi-parallel regions the particles can move more easily but with lower energy. With the above, the geometrical features, the velocity and the magnetic field strength of the shock, even the shock age are important factors for the acceleration of the particles (e.g., \citealt{2016LRSP...13....3D} and references therein).

\begin{figure}
\centering
\includegraphics[width=0.4\textwidth]{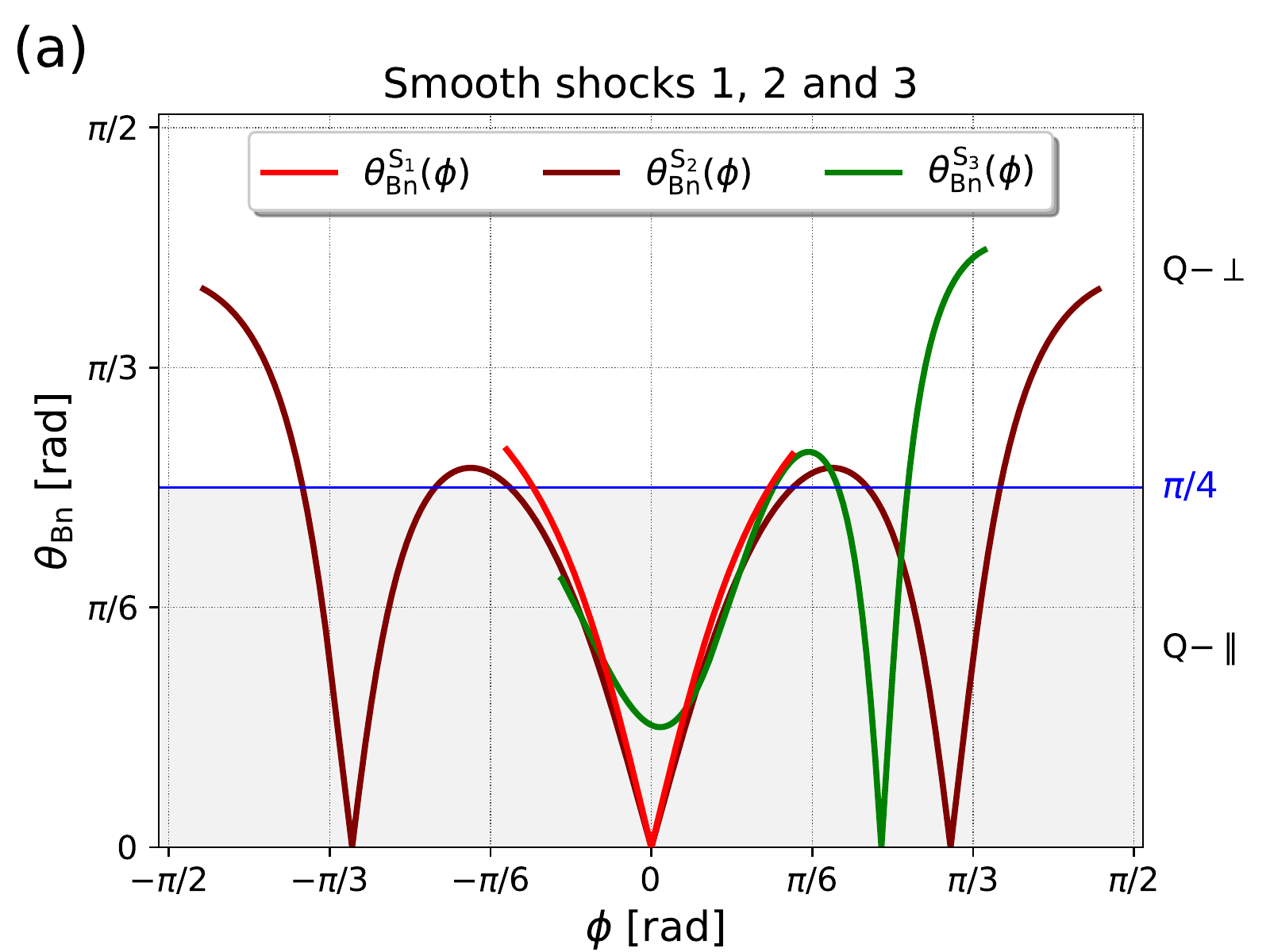} \\
\includegraphics[width=0.4\textwidth]{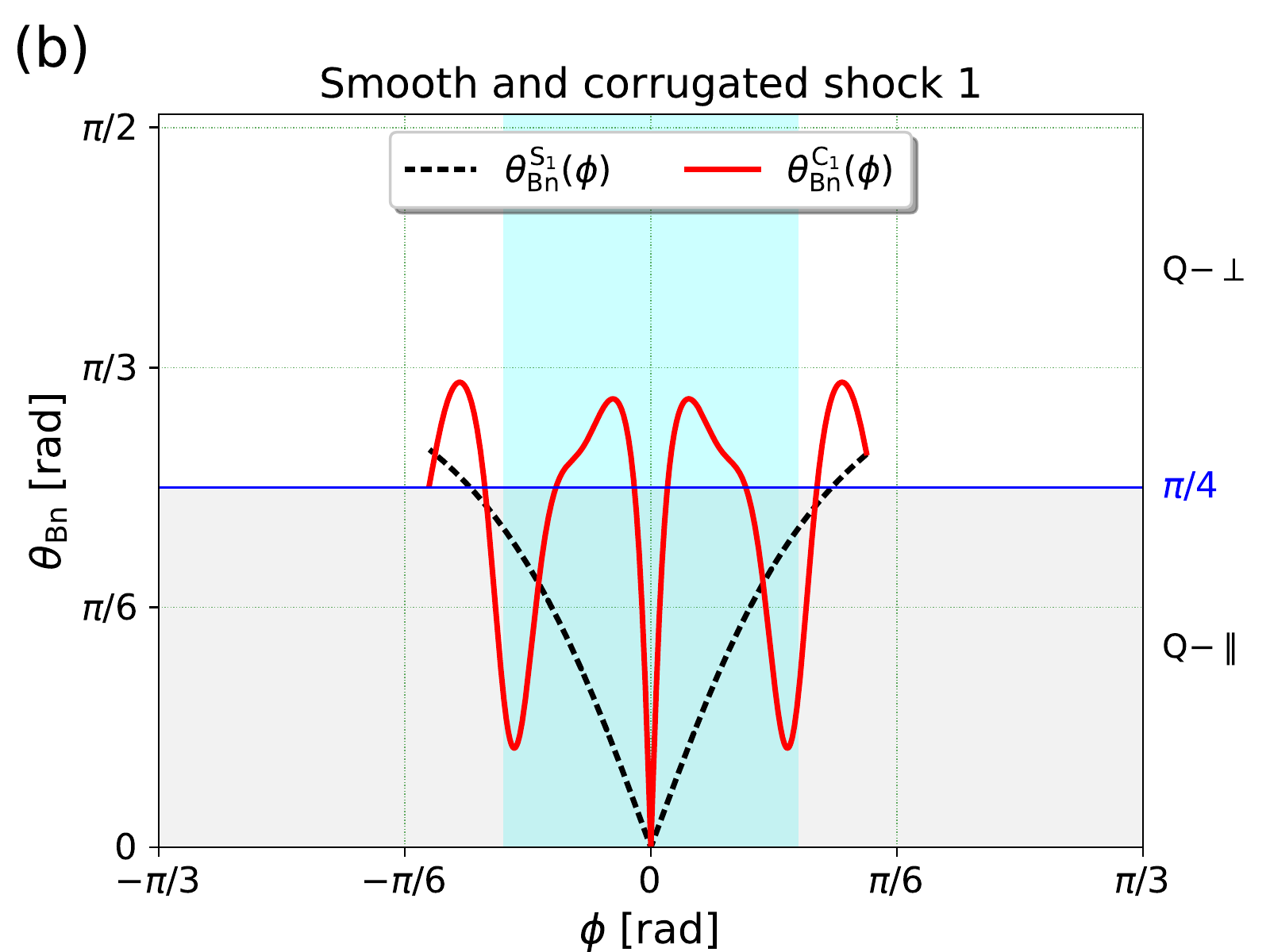} \\
\includegraphics[width=0.4\textwidth]{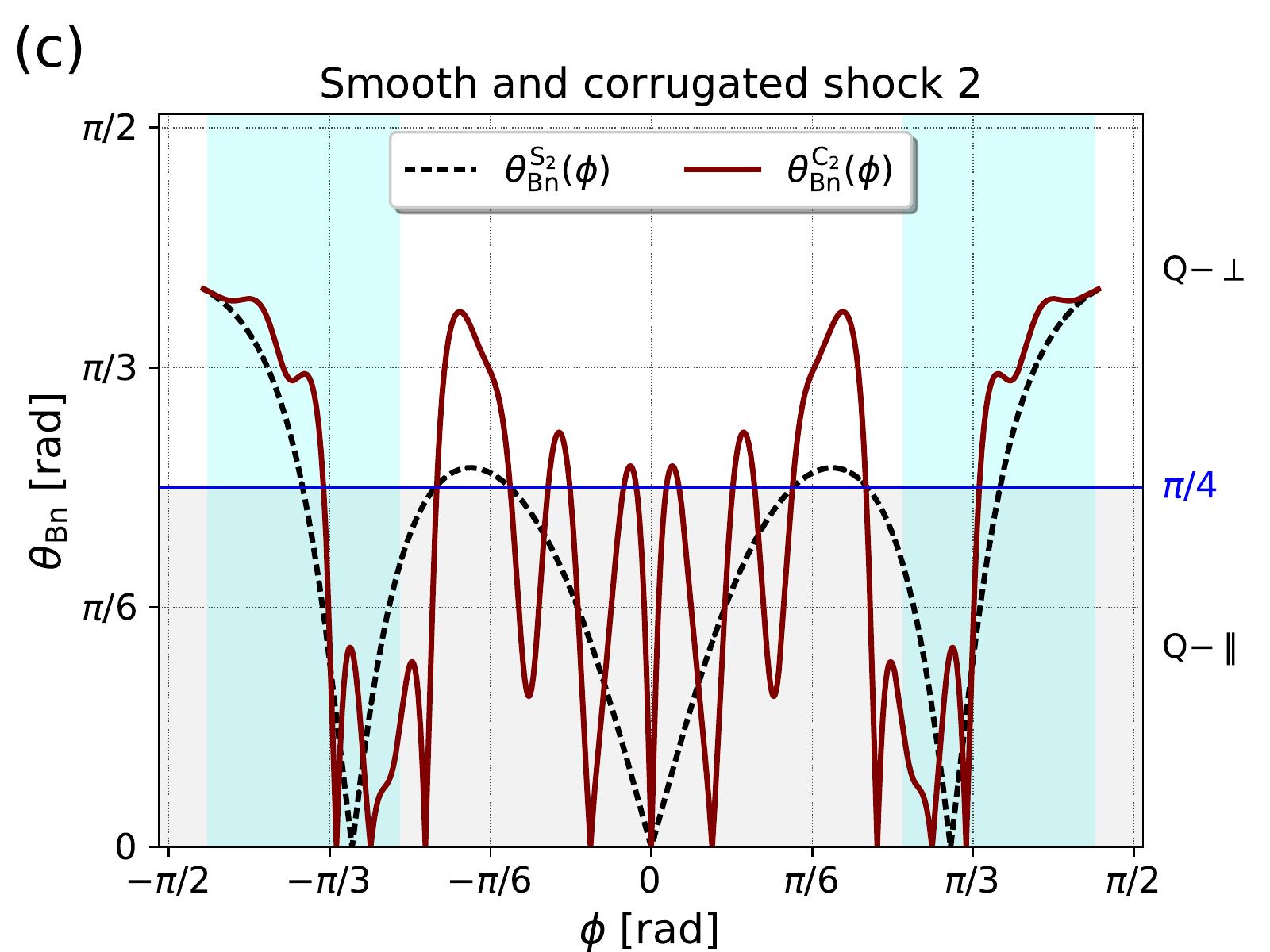} \\
\includegraphics[width=0.4\textwidth]{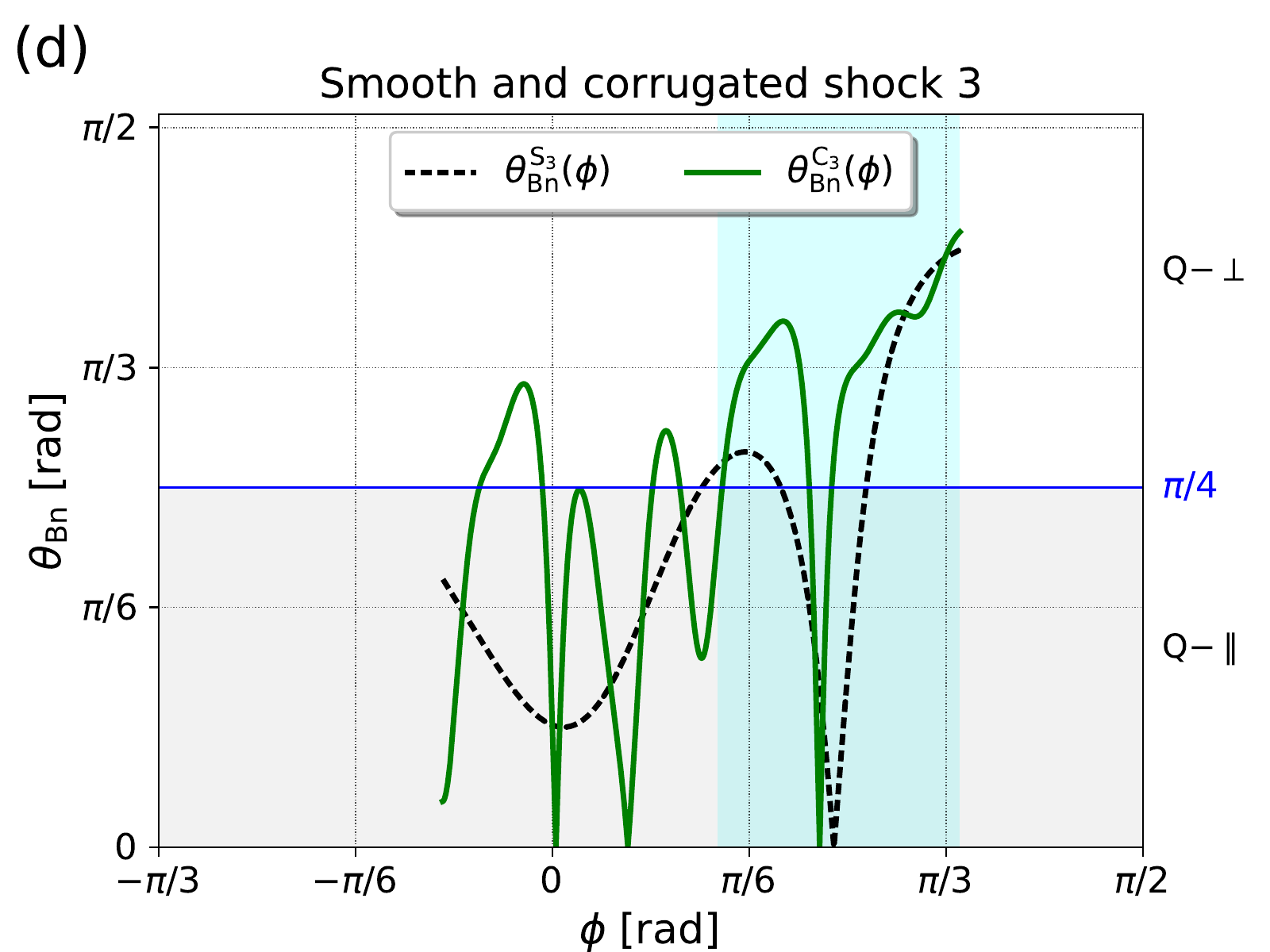}
\caption{Plots of shock normal angles for smooth, $\thetabn^{\rm S_m}(\phi)$, and corrugated $\thetabn^{\rm C_m}(\phi)$ shocks, with $m=1,2,3$, see Figure~\ref{plot-shocks}. Panel (a) shows the $\thetabn$ values for the three smooth shocks, and panels (b), (c) and (d) show separately the $\thetabn^{\rm S_m}(\phi)$ (black dotted line), and $\thetabn^{\rm C_m}(\phi)$ (color continuous line) values for each one of the shocks. The color plots are associated to the color plots feature in Figures~\ref{fig-cmes} and~\ref{plot-shocks}, i.e., the shocks 1, 2, 3 in red, maroon and green, respectively. The quasi-parallel (Q{\ndash}$\parallel$, $0 \leq \thetabn \leq \pi/4$) and quasi-perpendicular (Q{\ndash}$\perp$, $\pi/4 \leq \thetabn \leq \pi/2$) regions around the shocks are indicated by gray and white background shadow colors separated by blue line of $\theta= \pi/4$ rad. The cyan shadows in the plots indicate the initial supercritical regions assumed from Figure~\ref{plot-shocks}. The results between smooth and corrugated shocks are illustrated in Figure~\ref{fig-plane_shock} by distributions of filamentary structures in the upstream regions for quasi-parallel $\thetabn$ values.}\label{plot-thetabn_angles}
\end{figure}

\section{Discussion and conclusions}\label{sec-discuss}

The CME-driven shock formation in the low corona is a phenomenon that involves several physical processes as particle acceleration \citep[e.g.,][]{2005ApJ...622.1225M, 2013ApJ...778...43K}. Recently, some works show the relevance of including realistic features of the solar corona and CMEs in studies of SEPs \citep[e.g.,][]{2015ApJ...810...97S, 2017ApJ...836...36P, 2017ApJ...851...38K}. In this work, we discuss the predisposition to acceleration of particles in shocks with wave-like features imposed from ubiquitous disturbances of the solar corona, SW, and the corrugated CME piston. These wavy shocks are known as corrugated shocks \citep[e.g.,][]{1964PhFl....7..700G, 1987flme.book.....L}. Our work is motivated from observations showed in \cite{2015ApJ...812..119S}, where the authors show irregular shock front between 2 to 12 $\rs$ and 110$^{\circ}$ angular width. Similar shocks front were evidenced in \cite{2011ApJ...739L..64B}, \cite{2013JAdR....4..287B}, and \cite{2014ApJ...784..102B} at $\sim 2.5 \, \rs$ distances, in CME event of 1999 June 11. In this paper we calculate the shock normal angles, $\thetabn$, in order to interpret the physical process at the shock front \citep[e.g.,][]{2013pcss.book.....B}. With $\thetabn$, we identify the quasi-parallel ($0 \leq \thetabn \leq \pi/4$) and the quasi-perpendicular ($ \pi/4 \leq \thetabn \leq \pi/2$) regions, associated to particle acceleration. We do not study the evolutionary process in the shock, in contrast, we analyze the shock at $\sim 3.0 \, \rs$, in order to understand the predisposition of injection of particle through $\thetabn$, during the early stages of the shock where the particle acceleration is high \citep[e.g.,][]{2016LRSP...13....3D}.

In this paper, we analyze three different CME-driven shock morphologies from \cite{2009ApJ...693..267O}. In Figure~\ref{fig-cmes}, we show CME 1, CME 2 and CME 3, with respective shocks located in different latitudes i.e., CME nose (CME 1), both CMEs flanks (CME 2), and superior CME flank (CME 3). We define the smooth shocks $S_m(\phi)$, with $ m = 1, 2, 3 $, Equation~\eqref{eq-shock1},~\eqref{eq-shock2} and~\eqref{eq-shock3} through polar Gaussian plots as functions of the polar angular coordinate, $\phi$, see Figure~\ref{plot-shocks}(a). The corrugated shock $C_m(\phi)$, Equation~\eqref{eq-shock_und}, showed in Figure~\ref{plot-shocks}(b), is defined from $S_m(\phi)$ in addition with a complementary function $k(\phi)$, Equation~\eqref{eq-kphi} shown in Figure~\ref{plot-kphi}. Our study is focused in $\sim 3.0$, according to the shock formation, i.e., $\sim 1.5 \, \rs$ \citep[e.g.,][]{2011ApJ...738..160M, 2014A&A...564A..47Z, 2016ApJ...833..216G}, and SEPs onset $\sim 3.0 \, \rs$ \citep[e.g.,][]{2009ApJ...706..844R, 2012SSRv..171...23G}, for this reason we assume an amplitude less than 30\% $\rs$ in $k(\phi)$. We consider that wavy features in shocks are consequence of the disturbances from SW medium and CME deflection ($k_1(\phi)$, Equation~\eqref{eq-kphi_1}), irregular CME expansions or initial configuration of the CME ($k_2(\phi)$, Equation~\eqref{eq-kphi_2}), and minor disturbances from the solar corona, e.g., due to the fluctuation of density and Alfv\'en speed ($k_3(\phi)$, Equation~\eqref{eq-kphi_3}).

For the case of the smooth shocks, we find constraints between the shocks surfaces and their $\thetabn$ angles. Each shock shape can be interpreted like a sum of convex an concave contributions, i.e., like a compound of the profile of shock 1, $S_1(\phi)$, Equation~\ref{eq-shock1}. Consequently its $\thetabn$ profile may be a combination of $\thetabn^{\rm S_1}(\phi)$. For a situation similar to the shock decentralization, possibly due to deflection of CME or effect of external factors like coronal hole \citep[e.g.,][]{2012ApJ...755...43W}, its $\thetabn$ profile can be written as the shock normal angles of shock non-decentralized, i.e., the maximal and minimal $\thetabn$ values may be preserved. Therefore, the morphology of the $S_1(\phi)$ turns into a crucial element for interpreting others complex shocks like shock 2, $S_2(\phi)$ Equation~\eqref{eq-shock2}, and shock 3, $S_3(\phi)$ Equation~\eqref{eq-shock3}. The main diagnosis in the corrugated shocks is the fast oscillation of $\thetabn$ through the polar angle $\phi$. Figure~\ref{plot-thetabn_angles} shows how the $\thetabn^{\rm C_m}(\phi)$ changes drastically compared to the smooth shocks. 
The different amplitude and frequency in the components of $k(\phi)$ i.e., $k_1(\phi)$, $k_2(\phi)$ and $k_3(\phi)$, modify the $\thetabn$ profile of the smooth shocks. From Figure~\ref{plot-kphi}, the $k_2(\phi)$ function may be the most relevant source of disturbances in the shock. This function preserves intermediate amplitude and frequency values that $k_1(\phi)$ and $k_3(\phi)$, besides that represent the irregular shock features due to initial CMEs magnetic configurations and imbalance of magnetic pressures between CME and corona. Our results of $\thetabn^{\rm C_m}(\phi)$ do not evidence constraints similar to the smooth shock case. The $\thetabn$ oscillation from quasi-parallel to quasi-perpendicular values along the corrugated shocks evidence particles accelerated with different energy.  \cite{2005ApJ...624..765G} show that in quasi-perpendicular regions the particles are accelerated to high energies due to different coefficients of diffusion in these regions (e.g., \citealt{2016LRSP...13....3D} and references therein). With results of corrugated $\thetabn$ we consider that role of disturbances from solar corona, SW, and irregularities in the CME, may be relevant factor that define the energy of particles accelerated in shock front.

In this work, we do not study the evolution of the shock morphology, but \cite{2015ApJ...812..119S} observations, suggest that shock morphology evolution preserve the initial irregularities. In this way, the initial profile is a starting point for analyzing the shock evolution. Moreover, if the most notable shock attributes are preserved, the quasi-parallel and quasi-perpendicular regions will also be conserved. With this hypothesis, particle acceleration region in the shock may be maintained in the shock front, possibly with the exception of the angular width. Our idea of corrugated shocks evidence that the injection of particles through the shock may be complex due to the dependence of injection velocity with $\thetabn$ \citep[e.g.,][]{2012AdSpR..49.1067L, 2013pcss.book.....B}. In this way, the disturbances from coronal medium and corrugated CME piston may modify completely the SEPs production along the shock.

We understand that the physics of the corrugated CME-driven shocks is more complex than described here. The disturbances studied in this paper that depend on several factors, as ubiquitous fluctuations of the CME and solar corona properties, or even solar cycle phase. The corrugations features in CME-driven shocks are aleatory phenomena. Therefore, the perturbations on CME piston and shock, may be more complex than here showed. In this way, the future work, simulations or observations, may preserve some differences with our investigations. We considered that corrugated CME-driven shock may be a relevant element in order to understand the population and transport of SEPs through interplanetary space. These wave-like features on the shocks, may also be the started point for new research topics in the sheath region, e.g., like the downstream-jets and secondary shock, similar to the detected in the magnetosheath \citep{2009PhRvL.103x5001H, 2013JGRA..118.7237H}; sheath evolution process, e.g., like density variations and flows, e.g., Kelvin{\ndash}Helmholtz instability \citep{2005ApJ...622.1225M}; and particle acceleration in sheath region in high density regions (pile-up regions in \citealt{2011ApJ...729..112D}) \citep[e.g.,][]{2013ApJ...778...43K}.

\section*{Acknowledgments}

This study was financed by the Coordena\c{c}\~ao de Aperfei\c{c}oamento de Pessoal de N\'ivel Superior, Brasil (CAPES), Finance Code 001, PROEX 3474/2014. V.J.P. and D.F.G. thank the Brazilian agency FAPESP (No. 2013/10559-5) for support.

\bibliographystyle{aasjournal}
\bibliography{paez.bib}

\end{document}